\documentclass[aps,twocolumn,pra,superscriptaddress,amsmath,showpacs,tightenlines]{revtex4-1}
\usepackage{amssymb}
\usepackage{amsmath}
\usepackage{graphicx}
\usepackage{subfigure}
\usepackage{natbib}
\usepackage{epsfig}
\usepackage{amsfonts}
\usepackage{mathrsfs}
\usepackage{xcolor}
\usepackage[toc,page,title,titletoc,header]{appendix}

\begin{document}

\title{Microwave degenerate parametric down-conversion with a single cyclic three-level system in circuit QED}

\author{Z. H. \surname{Wang}}
\affiliation{Beijing Computational Science Research Center, Beijing 100084,
China}

\author{C. P. \surname{Sun}}
\affiliation{Beijing Computational Science Research Center, Beijing 100084, China}
\affiliation{Synergetic Innovation Center of Quantum Information and Quantum Physics,
University of Science and Technology of China, Hefei 230026, China}
\author{Yong \surname{Li}}
\email{liyong@csrc.ac.cn}
\affiliation{Beijing Computational Science Research Center, Beijing 100084, China}
\affiliation{Synergetic Innovation Center of Quantum Information and Quantum Physics,
University of Science and Technology of China, Hefei 230026, China}

\begin{abstract}
With the assistance of a single cyclic three-level system, which can be realized by a superconducting flux qubit, we study theoretically the degenerate microwave parametric down-conversion (PDC) in a superconducting transmission line resonator with the fundamental and second harmonic modes involved. By adiabatically eliminating the excited states of the three-level system, we obtain an effective microwave PDC Hamiltonian for the two resonator modes  in such a circuit QED system. The corresponding PDC efficiency in our model can be much larger than that in the similar circuit QED system based on a single two-level superconducting qubit [K. Moon and S. M. Girvin, Phys. Rev. Lett. {\bf 95}, 140504 (2005)]. Furthermore, we consider the squeezing and bunching behavior of the fundamental mode resulting from the coherent drive to the second harmonic one.
\end{abstract}


\maketitle
\section{introduction}
Photonic parametric down-conversion (PDC) refers to the coherent generation of a pair of photons with lower-frequency via injecting a higher-frequency photon into the nonlinear medium~\cite{no}. The PDC together with three-wave mixing in atomic medium has been widely studied both theoretically~\cite{na,ca,rm,ba} and experimentally~\cite{Pavel,sw1,fs}. Via the PDC process, the squeezing state which is usually used in precise measurement~\cite{qa} can be generated and has been observed in optical cavity~\cite{re,lawu}.

In the early days, the degenerate~\cite{na} and non-degenerate~\cite{ca} PDC processes have been studied in cyclic three-level atomic system where any two of the three levels can be coupled via electric dipole transition. In general, the cyclic three-level structure does not exist in natural atoms due to the rules of electric dipole transitions. The key point to form a cyclic atomic structure in Refs.~\cite{na,ca} is to use a sufficiently strong external field to break the symmetry of the system.

Recently, it has been found that such an electrical-dipole-transition based cyclic three-level (also called $\Delta$-type) structure can be formed in the system of three-level flux qubit~\cite{Yuxiprl} by adjusting the bias magnetic flux threaded through the loop formed by three Josephson junctions. Besides the flux qubit, the cyclic three-level structures also exist in chiral molecular systems~\cite{kral,kral2}, and are used to separate the chiral molecules with different chiralities by generalized Stern-Gerlach effect~\cite{yongli}. Based on the cyclic transitions, it is convenient to use $\Delta$-type three-level systems to generate single microwave photons~\cite{youprb,yuxi}, produce microwave amplification without population inversion~\cite{wz,joo} and serve as a single-photon quantum router~\cite{Lan13}.

On the other hand, there have been great progresses in simulating quantum optics phenomenon in circuit QED system~\cite{you,im,jli}, where the superconducting qubit (e.g. charge, phase or flux qubit) serves as a two-level or three-level ``artificial atom'' interacting with the microwave superconducting transmission line resonator. In the circuit QED system, the energy structure of qubits can be easily controlled by tuning the external conditions such as currents, voltages, and electromagnetic fields. The strong couplings between artificial atoms and superconducting resonators have also been realized experimentally
~\cite{Wallraff,Niemczyk,Ichi}. Therefore, people have proposed to realize the PDC process in the circuit QED system where two of microwave modes in the superconducting transmission line resonator are coupled to the two-level superconducting qubit(s)~\cite{girvin,long,KI}.

Based on the above achievements, we consider in this paper the microwave PDC and generation of the squeezed state in a circuit QED system consisting of a single cyclic three-level superconducting flux qubit and a two-mode transmission line resonator. In the case that the detunings between the qubit and the two resonator modes are much larger than their coupling strengths, we can eliminate adiabatically the degrees of freedom of the qubit and derive the effective coupling between the two modes in the resonator by Fr\"{o}lich-Nakajima transformation~\cite{H. Frohlich} (also called Schrieffer-Wolff transformation~\cite{JR1,JR2}). The effective Hamiltonian has the similar form as that of the degenerate parametric oscillator~\cite{DF,mj}, which supports the PDC process and the generation of the squeezing field.

In our proposal, the efficiency of the microwave PDC is inversely proportional to the detuning between the fundamental mode and the qubit, and therefore much larger than that in the system of two-level qubit interacting with the superconducting transmission line resonator, in which the efficiency is inversely proportional to the frequency of the qubit's transition~\cite{girvin}. We further resonantly drive the second harmonic mode and discuss the squeezing and bunching behavior of the fundamental mode by means of the mean field approach~\cite{wu}.

The rest of the paper is organized as follows. In Sec.~\ref{s2}, we illustrate our model and derive the effective PDC Hamiltonian via adiabatically eliminating the degrees of freedom of the qubit. In Sec.~\ref{steady}, we investigate the squeezing and bunching behavior of the fundamental mode in superconducting transmission line resonator by means of the Langevin equations. In Sec.~\ref{s4}, we give some brief conclusions.

\section{The model and parametric down-conversion}\label{s2}

\begin{figure}[tbp]
\centering
\includegraphics[width=8cm]{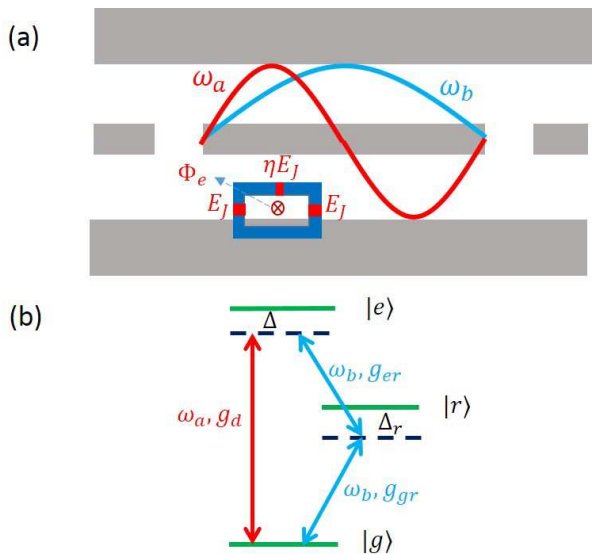}
\caption{(Color online) (a) The schematic diagram of the circuit QED consisting of a flux qubit and a two-mode microwave superconducting transmission line resonator. (b) The cyclic three-level structure of the flux qubit interacting with the two modes of the microwave resonator.}
\label{scheme}
\end{figure}

As shown in Fig.~\ref{scheme}(a), we consider a circuit QED system with a three-level superconducting flux qubit interacting with a two-mode transmission line resonator. The qubit is composed of a superconducting loop with three Josephson junctions. Two of the junctions have equal Josephson energies $E_J$, while the third one has $\eta E_J$ $(1/2<\eta<1)$.  The loop is threaded by an external magnetic flux $\Phi_e$.  When the threading flux satisfies $\Phi_e=0.5\Phi_0$, where $\Phi_0=\pi\hbar c/e$ is the flux quanta with $e$ the electronic charge and $c$ the speed of light in vacuum, the potential and the energy eigen-states of the flux qubit have fixed parities so that the electrical-dipole transitions between the states with same parity are forbidden. In this case, the lowest three states of the flux qubit can just form a cascade three-level structure. However, when $\Phi_e\neq 0.5\Phi_0$, the symmetry of the potential is broken, the transitions between arbitrary two states are possible. Thus the flux qubit can form a cyclic $\Delta$-type energy level configuration~\cite{Yuxiprl}, allowing for the coexistence of one- and two-photon processes~\cite{deppe}.

We consider that two of the modes (i.e., the fundamental and second harmonic modes) in the superconducting transmission line resonator are involved. The second harmonic mode of frequency $\omega_a$ couples the transition between the ground state $|g\rangle$ and the second excited state $|e\rangle$ of the qubit with the coupling strength $g_d$. The fundamental mode of frequency $\omega_b$ ($=\omega_a/2$) couples the $|g\rangle\leftrightarrow|r\rangle$ (with $|r\rangle$ the first excited state of the flux qubit) and $|e\rangle\leftrightarrow|r\rangle$ transitions simultaneously, with the coupling strengths $g_{\rm{er}}$ and $g_{\rm{gr}}$, respectively [as shown in Fig.~\ref{scheme}(b)].

The Hamiltonian of the circuit QED system is written as $H=H_0+H_I$, where (hereafter we set $\hbar=1$)
\begin{equation}
H_0=\omega_{e}|e\rangle\langle e|+\omega_{r}|r\rangle\langle r|+\omega_a a^{\dagger}a+\omega_{b}b^{\dagger}b
\label{h0}
\end{equation}
is the free Hamiltonian of the three-level qubit and the microwave modes in the resonator. Here, $\omega_r$ and $\omega_e$ are the energies of the
first excited state $|r\rangle$, and the second excited state $|e\rangle$, respectively. As a reference, we have set the energy of the ground state $|g\rangle$ as $\omega_g=0$. $a^{\dagger}$ ($b^{\dagger}$) is the creation operator for the second harmonic (fundamental) modes. The interactions between the resonator modes and the three-level qubit are described by
\begin{equation}
H_I=g_{d}a|e\rangle\langle g|+g_{\rm{gr}}b|r\rangle\langle g|
+g_{\rm{er}}b|e\rangle\langle r|+h.c.,
\label{hi}
\end{equation}
where we have used the rotating wave approximation. In order to obtain the effective coupling between the two modes in the resonator, we should eliminate the degrees of freedom of the qubit. Here, we adopt the Fr\"{o}lich-Nakajima transformation, which is a canonical transformation widely used in condensed matter physics~\cite{H. Frohlich} and quantum optics~\cite{HB,ly}, to eliminate the variables of the qubit. To this end, we first define the detunings as [shown in Fig.~\ref{scheme}(b)],
\begin{eqnarray}
\Delta&:=&\omega_e-\omega_a=\omega_e-2\omega_b,\\
\Delta_r&:=&\omega_r-\omega_b.
\end{eqnarray}
In the regime of large detunings $\{ |\Delta|, |\Delta_r| \}\gg \{|g_d|, |g_{\rm{er}}|, |g_{\rm{gr}}| \}$, the effective coupling between the two modes in the resonator can be obtained by introducing the unitary transformation $\mathcal{H}=\exp(-S)H\exp(S)$, where
\begin{equation}
S=\frac{g_{d}}{\Delta}a^{\dagger}|g\rangle\langle e|+\frac{g_{\rm{er}}}{\Delta-\Delta_{r}}b^{\dagger}|r\rangle\langle e| +\frac{g_{\rm{gr}}}{\Delta_{r}}b^{\dagger}|g\rangle\langle r|-h.c.
\end{equation}
Here we also have assumed $|\Delta-\Delta_r | \gg |g_{\rm{er}}|$ in our consideration.
\begin{figure}[tbp]
\centering
\includegraphics[width=8cm]{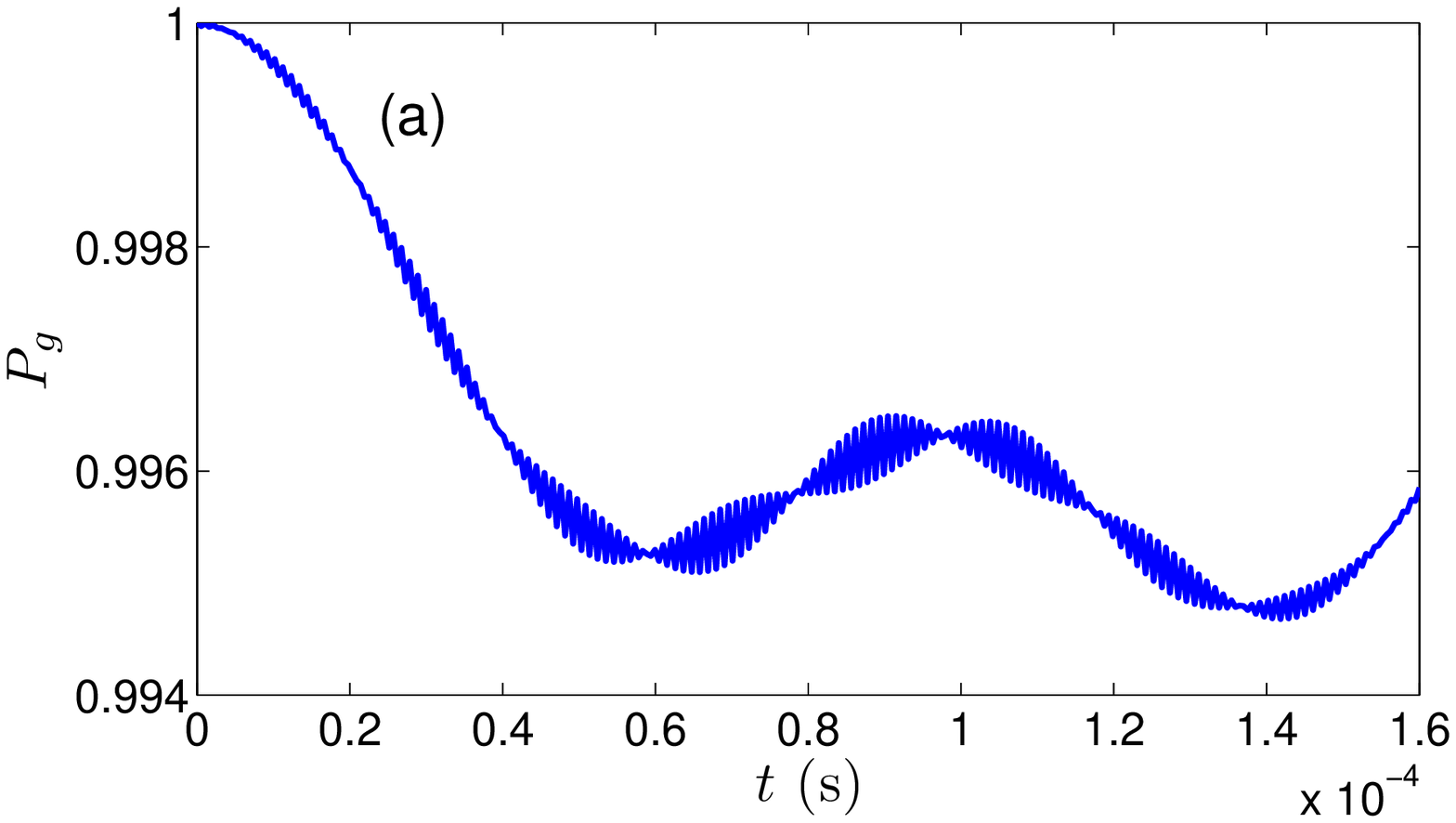}
\includegraphics[width=8cm]{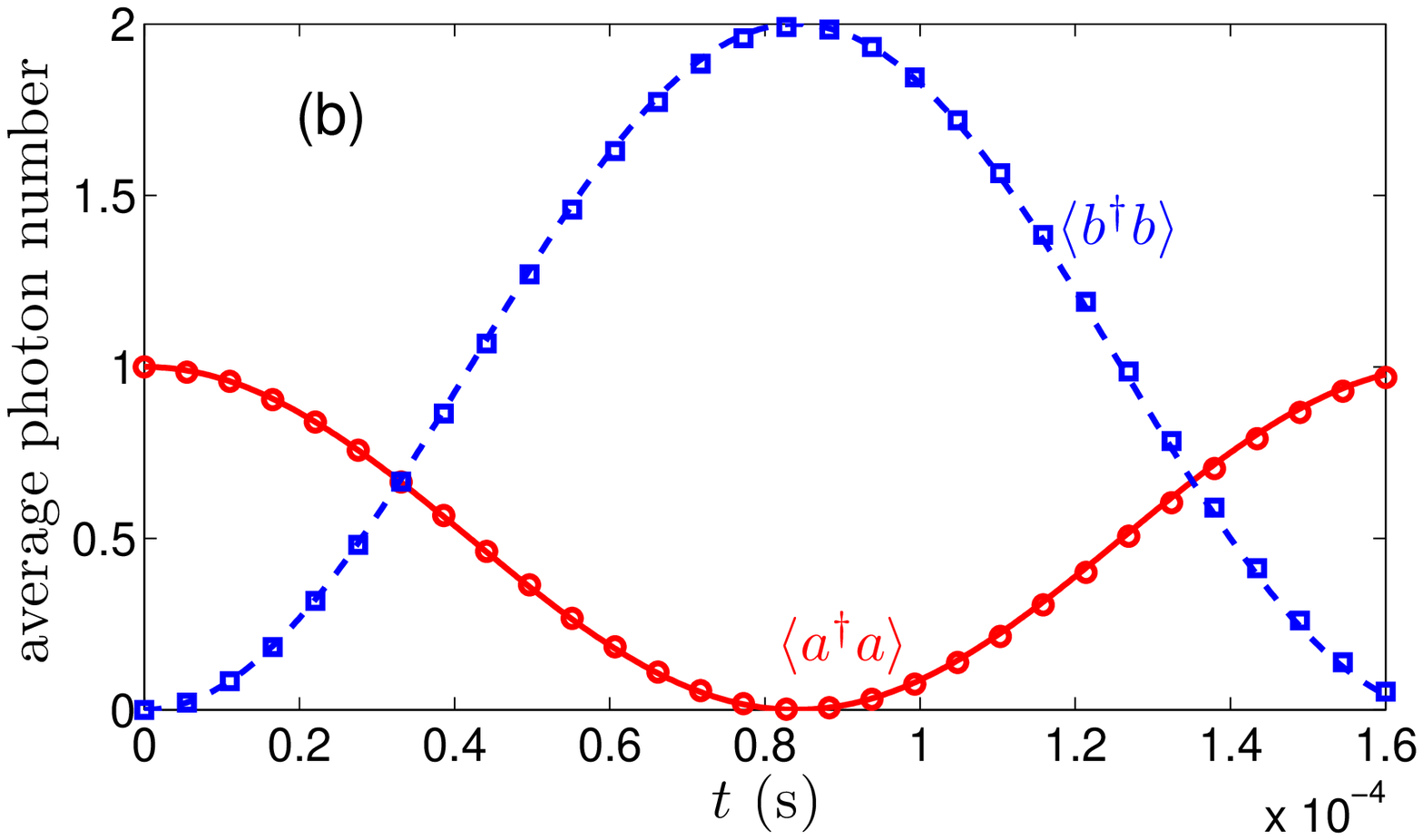}
\caption{(Color online) (a) The probability for the qubit in the ground state. (b) The average photon numbers of the second harmonic mode (red solid line) and fundamental mode (blue dashed line) in the resonator. The empty circles and rectangles are the corresponding numerical results. The parameters are set as $ \omega_{a}/2\pi=2\omega_{b}/2\pi=5.5$ GHz, $\Delta=2\Delta_r=\omega_a/10$ and $|g_d|/2\pi=20$ MHz, $|g_{\rm{gr}}|/2\pi=|g_{\rm{er}}|/2\pi=10$ MHz. Under these parameters, the effective coupling strength is $\chi/2\pi\approx26$ kHz, and the modification frequency of the second
harmonic mode is $\omega_{\rm{eff}}\approx5.5$ GHz. We assume the system is prepared in the state $|\psi(0)\rangle=|g;1;0\rangle$ initially.}
\label{numerical}
\end{figure}

Since we focus on considering the situation of large detunings, the qubit populated in the initial ground state $|g\rangle$ will not exchange photons with the resonator and will be remained in the ground state. Neglecting the high-frequency terms and the virtual photon induced modification to the energy of the qubit's excited states, one can obtain the following effective Hamiltonian
\begin{equation}
 \mathcal{H}=H_{\rm{eff}}\otimes|g\rangle\langle g|.
\end{equation}
Here, up to the third order of the interaction, the effective Hamiltonian for the two resonator modes is given as
\begin{eqnarray}
H_{\rm{eff}}&=&\langle g|\mathcal{H}|g\rangle\nonumber \\&\approx&\langle g|(H_{0}+\frac{1}{2}[H_{I},S]+\frac{1}{3}[[H_{I},S],S])|g\rangle \nonumber \\&=&(\omega_a-\frac{|g_{d}|^{2}}{\Delta})a^{\dagger}a+(\omega_b-\frac{|g_{\rm{gr}}|^{2}}{\Delta_{r}})b^{\dagger}b\nonumber \\&&+\left(\frac{g_{d}g_{\rm{er}}g_{\rm{gr}}}{\Delta_{r}\Delta} a^{\dagger}b^{2}+h.c.\right),
\label{eff}
\end{eqnarray}
where $-|g_d|^2/\Delta$ and $-|g_{\rm{gr}}|^2/\Delta_r$ are the frequency shifts of the second harmonic and fundamental modes due to the largely-detuned couplings to the qubit. Usually, these shifts are negligibly small compared with the corresponding resonant frequencies.

In the case of~\cite{note}
\begin{equation}
\omega_{a}-\frac{|g_{d}|^{2}}{\Delta}=2(\omega_{b}-\frac{|g_{\rm{gr}}|^{2}}{\Delta_{r}})=:\omega_{\rm{eff}},
\label{cond}
\end{equation}
the effective Hamiltonian becomes
\begin{equation}
H_{\rm{eff}}=\omega_{\rm{eff}} a^{\dagger}a+\frac{\omega_{\rm{eff}}}{2}b^{\dagger}b+\frac{\chi}{2}(e^{i\varphi}a^{\dagger}b^{2}+e^{-i\varphi}ab^{\dagger2}),
\end{equation}
where
\begin{equation}
\chi=\frac{2|g_{d}g_{\rm{er}}g_{\rm{gr}}|}{\Delta_{r}\Delta}
\end{equation}
is the effective coupling strength between the two modes in the resonator, and $\varphi$ is the global phase contributed from the three couplings between the resonator and
the qubit.

In what follows, we will choose the parameters as~\cite{girvin} $\omega_{a}/2\pi=2\omega_{b}/2\pi=5.5$ GHz, $\Delta=2\Delta_r=\omega_a/10$, and $|g_d|/2\pi=20$ MHz, $|g_{\rm{gr}}|/2\pi=|g_{\rm{er}}|/2\pi=10$ MHz. Under these parameters, the effective coupling strength is $\chi/2\pi\approx26$ kHz and the effective
modification frequency of the second harmonic mode is $\omega_{\rm{eff}}\approx5.5$ GHz . In order to verify the validity of the adiabatic elimination, we illustrate the time evolution of the system in Fig.~\ref{numerical}, assuming that the system is initially prepared in the state $|g;1;0\rangle\equiv|g\rangle_q\otimes|1\rangle_a\otimes|0\rangle_b$, which means that the flux qubit is in its ground state, and the second harmonic (fundamental) mode in the resonator is in the Fock state $|1\rangle$ ($|0\rangle$). In Fig.~\ref{numerical}(a), we plot the probability for the flux qubit remained in the ground state $|g\rangle$ during the time evolution governed by the Hamiltonians~(\ref{h0},\ref{hi}) without considering the dissipation of the resonator modes and the flux qubit. The fact that the probability even surpasses $0.99$ means it is reasonable to assume that the qubit is always populated in the ground state. Moreover, it shows obvious Rabi oscillation between the states $|g;1;0\rangle$ and $|g;0;2\rangle$ in Fig.~\ref{numerical}(b), where we plot the average photon numbers as a function of the evolution time $t$. We also observe from Fig.~\ref{numerical}(b) that our results based on the effective Hamiltonian~(\ref{eff}) (represented by the dashed and solid lines) coincide with the direct numerical results (represented by the empty rectangles and circles) based on the original Hamiltonian in Eqs.~(\ref{h0},\ref{hi}). This further grantees the validity of the method of adiabatic elimination we used here.

The effective Hamiltonian in Eq.~(\ref{eff}) demonstrates a degenerate PDC mechanism via nonlinear three-wave mixing. Actually, a similar process has also been investigated in the system of two-level qubit interacting with superconducting transmission line resonator in Ref.~\cite{girvin}. However, in Ref.~\cite{girvin}, the PDC efficiency $\chi$ is inversely proportional to $\omega_a$, which is always in the order of GHz in current experiments~\cite{you,Buluta}. On the contrary, in our scheme of cyclic three-level qubit, the PDC efficiency is inversely proportional to the detuning $\Delta_r$, which can be much smaller than $\omega_a$. Therefore, the PDC efficiency has been enlarged significantly in our scheme.

\section{squeezing and photonic correlation}
\label{steady}

Now, we will study the optical character of the effective PDC system in the steady state at the presence of the driving and dissipation simultaneously. Here, we focus on the situation that the second harmonic mode is resonantly driven by an external field and the action of the driving field is described by
\begin{equation}
H_{\rm{drive}}=i\epsilon(a^{\dagger}e^{-i\omega_{\rm{eff}}t}-ae^{i\omega_{\rm{eff}}t}),
\end{equation}
where $\epsilon$ is the strength of the driving field and assumed to be real.

In the rotating frame with respect to $U(t)=\exp[i\omega_{\rm{eff}} (a^{\dagger}a+ b^{\dagger}b/2)t]$, the Hamiltonian of the system under the driving field becomes
\begin{eqnarray}
\tilde{H} & = & U(t)(H_{\rm{eff} }+H_{\rm {drive}})U^{\dagger}(t)+i\frac{\partial U(t)}{\partial t}U^{\dagger}(t)\nonumber \\
 & = & i\epsilon(a^{\dagger}-a)+\frac{i\chi}{2}(ab^{\dagger2}-a^{\dagger}b^{2}).
 \label{pdc}
\end{eqnarray}
Here, we have set the global phase as $\varphi=-\pi/2$.

Based on the above standard PDC Hamiltonian~(\ref{pdc}), the Langevin equations
for the operators $a$ and $b$ are
\begin{eqnarray}
\frac{d}{dt}a & = & \epsilon-\frac{\chi}{2}b^{2}-\gamma_{a}a+\sqrt{2\gamma_{a}}a_{in},\label{eq:a}\\
\frac{d}{dt}b & = & \chi ab^{\dagger}-\gamma_{b}b+\sqrt{2\gamma_{b}}b_{in},\label{eq:b}
\end{eqnarray}
where the noise operators satisfy $\langle a_{in}(t)a_{in}^{\dagger}(t')\rangle=\langle b_{in}(t)b_{in}^{\dagger}(t')\rangle=\delta(t-t')$ since we have restricted our consideration at very low temperature such that the corresponding thermal photon numbers for the resonator modes are close to $0$. $\gamma_a$ and $\gamma_b$ are respectively the decay rates of the second harmonic and fundamental modes.

\begin{figure}[tbp]
\centering
\includegraphics[width=8cm]{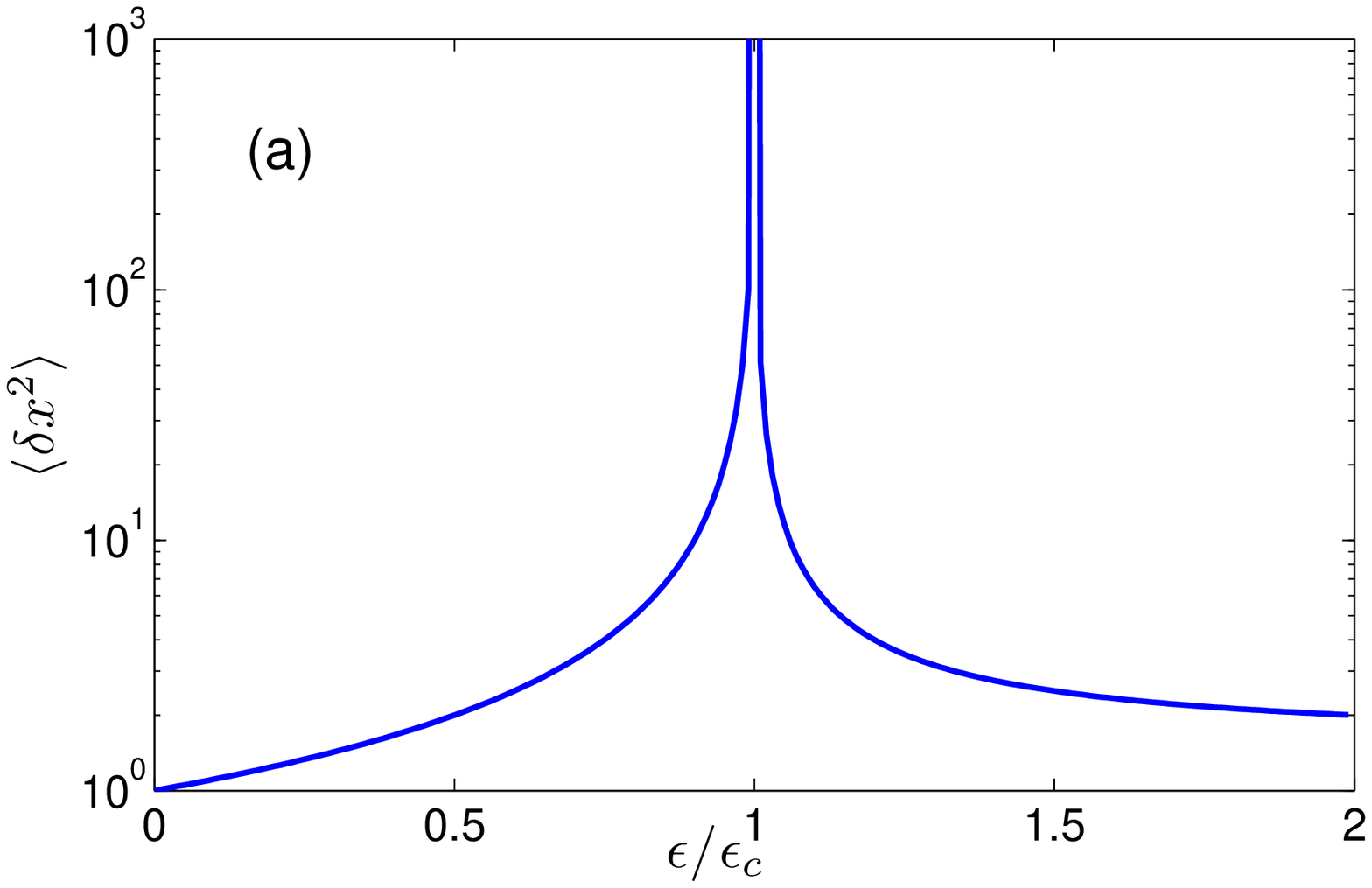}
\includegraphics[width=8cm]{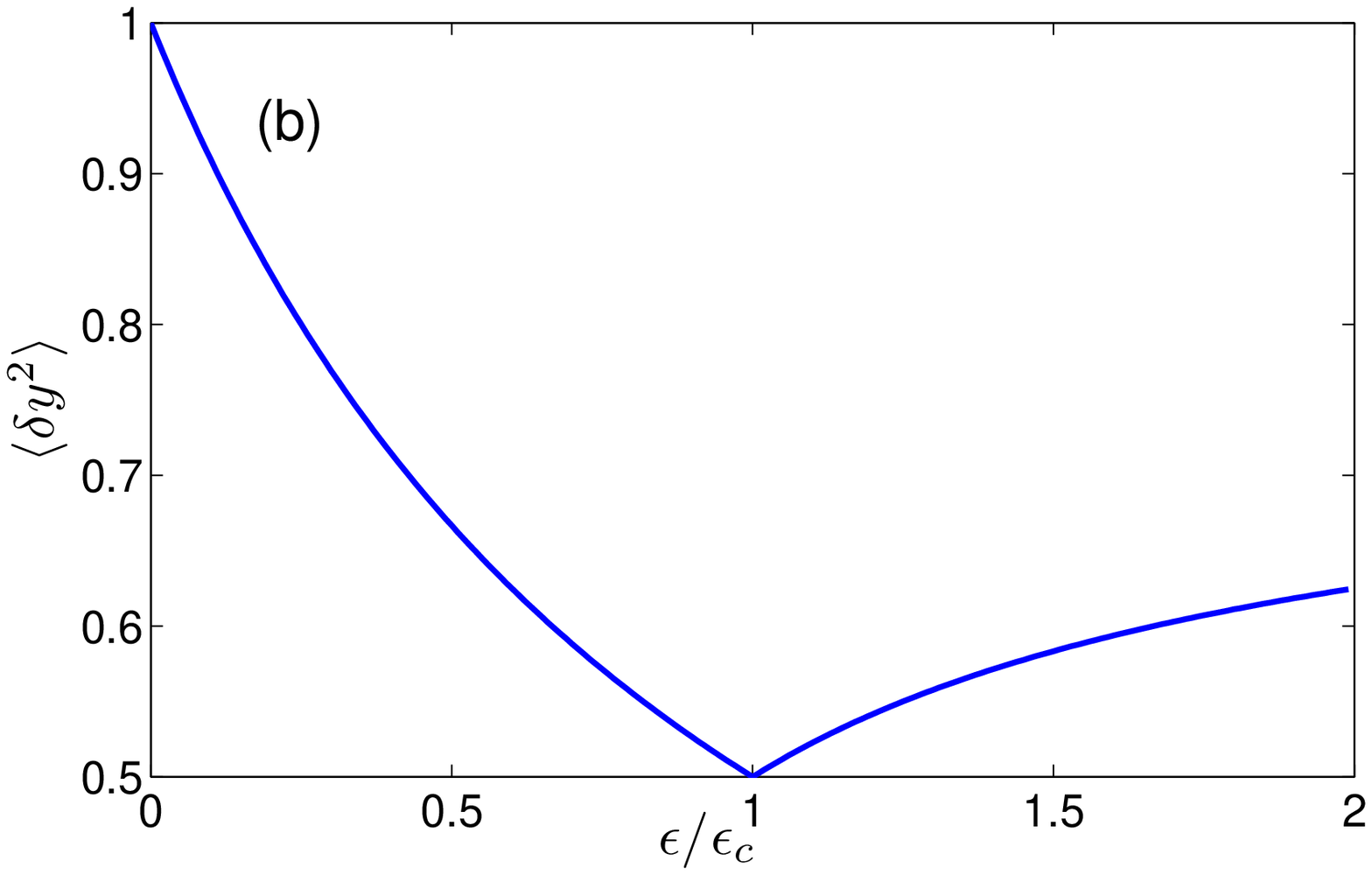}
\caption{(Color online) Variances of the quadratures (a) $\langle\delta x^2\rangle$ and (b) $\langle \delta y^2\rangle $ for
the fundamental mode as a function of the driving strength. The parameters are set as $\gamma_a/2\pi=11$ MHz, and $\gamma_b/2\pi=5.5$ MHz. For the other parameters, see Fig.~\ref{numerical}.}
\label{squeezing}
\end{figure}

In order to linearize the above equations, we define the operators as $a=\alpha+\delta a$, $b=\beta+\delta b$, where $\alpha$ ($\beta$) is the average value of the operator $a$ ($b$) in the steady state and $\delta a$ ($\delta b$) is its fluctuation. The solution of the average values are given by~\cite{wu,PDD}
\begin{equation}
\begin{cases}
\alpha=\epsilon/\gamma_{a},  \ \ \beta=0, & \epsilon\leq\epsilon_{c}\\
\alpha=\gamma_{b}/\chi, \ \ \beta=\pm\sqrt{\frac{2}{\chi}(\epsilon-\epsilon_{c})},& \epsilon>\epsilon_{c}
\end{cases}.
\end{equation}
Obviously, there is a phase transition at the critical driving strength (threshold) $\epsilon_c=\gamma_a\gamma_b/\chi$. In what follows, we only consider the positive branch for $\beta$ when the driving strength is above the threshold.

After neglecting the high-order terms of the fluctuations, we obtain the linearized quantum Langevin equations for the fluctuation operators $\delta a$ and $\delta b$ as
\begin{eqnarray}
\frac{d}{dt}\delta a&=&-\chi\beta\delta b-\gamma_{a}\delta a+\sqrt{2\gamma_{a}}a_{in},\\
\frac{d}{dt}\delta b&=&\chi\alpha\delta b^{\dagger}+\chi\beta^{*}\delta a-\gamma_{b}\delta b+\sqrt{2\gamma_{b}}b_{in},
\end{eqnarray}
which can be solved analytically by means of Fourier transformation.
Let us define the quadratures $\delta x(t)=\delta b(t)+\delta b^{\dagger}(t)$ and $\delta y(t)=-i[\delta b(t)-\delta b^{\dagger}(t)]$, the expressions of their variances are obtained as
\begin{equation}
\langle\delta x^2\rangle=\begin{cases}
\frac{\gamma_{a}\gamma_{b}}{\gamma_{a}\gamma_{b}-\epsilon\chi}, & \epsilon\leq\epsilon_{c}\\
1+\frac{\gamma_{b}}{\gamma_{a}}-\frac{\gamma_{a}\gamma_{b}}{2(\gamma_{a}\gamma_{b}-\epsilon\chi)}, & \epsilon>\epsilon_{c}
\end{cases},
\label{dx}
\end{equation}
and
\begin{equation}
\langle\delta y^2\rangle=\begin{cases}
\frac{\gamma_{a}\gamma_{b}}{\gamma_{a}\gamma_{b}+\epsilon\chi}, & \epsilon\leq\epsilon_{c}\\
1-\frac{\gamma_{a}^{2}\gamma_{b}}{(\gamma_{a}+2\gamma_{b})\epsilon\chi}, & \epsilon>\epsilon_{c}
\end{cases}.
\label{dy}
\end{equation}
We plot the variances as functions of the driving strength in Fig.~\ref{squeezing}. It can be observed from Eqs.~(\ref{dx},\ref{dy}) and Fig.~\ref{squeezing} that, $\langle\delta x^2\rangle$ is always larger than $1$, and $\langle\delta y^2\rangle$ is always smaller than $1$. In other words, when the second harmonic mode is coherently driven resonantly, the fundamental mode exhibits a squeezing effect. It can be observed in Fig.~\ref{squeezing}(a) that $\langle\delta x^2\rangle$ diverges when the driving strength
is close to the threshold. This implies that the linearization does not work well in this regime. Recently, the modification near the threshold has been made by regularized linearization approach, in which the steady values $\alpha$ and $\beta$ are determined self-consistently~\cite{carlos}. However, the results from the two approaches coincide with each other in the regime deviating from the threshold.

\begin{figure}
\centering
\includegraphics[width=8cm]{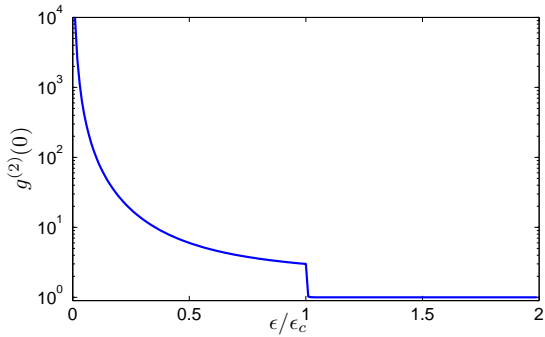}
\caption{The equal-time second-order correlation $g^{(2)}(0)$ as a function of the driving strength. The parameters are same as those in Fig.~\ref{squeezing}. }
\label{correlation}
\end{figure}

Furthermore, we can also investigate the statistic properties of the fundamental mode by calculating its equal-time second-order correlation. The second-order correlation of the fundamental mode is defined by
\begin{equation}
g^{(2)}(0)\equiv\frac{\langle b^{\dagger}(t)b^{\dagger}(t)b(t)b(t)\rangle}{\langle b^{\dagger}(t)b(t)\rangle^{2}},
\end{equation}
and the corresponding numerical results are shown in Fig.~\ref{correlation}. It is observed that $g^{(2)}(0)\gg 1$ when the driving strength is below the threshold, which implies a strong bunching character. As the increase of the driving strength, the correlation $g^{(2)}(0)$ decreases and approaches $1$ when $\epsilon>\epsilon_c$. It can be found that $g^{(2)}(0)$ experiences a sudden decrease at the threshold. This discontinuity can also be removed by the regularized linearization approach~\cite{carlos}.

\section{Conclusion}\label{s4}
In summary, we have studied the degenerate microwave PDC in the circuit QED system where a single cyclic three-level superconducting qubit couples to the fundamental and second harmonic modes in a transmission line resonator simultaneously. In the situation of large detunings, we adiabatically eliminate the degree of freedom of the qubit (that is, keeping the qubit in the ground state) and obtain the effective PDC Hamiltonian for the two microwave resonator modes. Within the available experimental parameters, we show that the method of the adiabatical elimination is reasonable by comparing the corresponding approximate analytical results with the direct numerical calculations. Compared with the scheme in which the two-mode resonator couples to a single two-level qubit~\cite{girvin}, the PDC efficiency in our model is dramatically enhanced with a single cyclic three-level flux qubit, which can be realized and tuned more easily in experiments. Based on the obtained effective Hamiltonian, we show that the coherent driving of the second harmonic mode will result in the squeezing and bunching effect of the fundamental mode. We hope that our proposal would open a way to generate the high-efficiency microwave PDC process in the system of circuit QED.

\begin{acknowledgments}
We thank C. N. Benlloch, S. W. Li and Y. X. Liu for their fruitful discussions.
This work is supported by the NSFC (under Grants No. 11174027 and No. 11121403) and the National 973 program (under Grants No. 2012CB922104 and No. 2014CB921403).

\end{acknowledgments}


\begin{thebibliography}{99}
\bibitem{note} Strictly speaking, the condition in Eq.~(\ref{cond}) can not be always satisfied. However, we can introduce another auxiliary two-level superconducting qubit, which is far-off detuned from one of the resonator modes, to provide an additional frequency shift to the mode such that Eq.~(\ref{cond}) is fullfilled. Alternativley,
    we can properly choose the detunings as in the maintext to fullfill Eq.~(\ref{cond}).
\end{thebibliography}
\end{document}